\documentclass[epj,twocolumn]{webofc}
\usepackage[varg]{txfonts}   % Web of Conferences font
\woctitle{CGS15}
\begin{document}
\title{Energy-dependence of skin-mode fraction
in $E1$ excitations of neutron-rich nuclei}
%
% subtitle is optionnal
%
%%%\subtitle{Do you have a subtitle?\\ If so, write it here}

\author{H.~Nakada\inst{1}\fnsep\thanks{\email{nakada@faculty.chiba-u.jp}} \and
        T.~Inakura\inst{1} \and
        H.~Sawai\inst{1}
        % etc.
}

\institute{Department of Physics, Graduate School of Science,
 Chiba University, Yayoi-cho 1-33, Inage, Chiba 263-8522, Japan
          }

\abstract{%
We have extensively investigated characters of the low-energy $E1$ strengths
in $N>Z$ nuclei,
by analyzing the transition densities obtained by the HF+RPA calculations
with several effective interactions.
Crossover behavior has been confirmed,
from the skin mode at low energy to the $pn$ mode at higher energy.
Decomposing the $E1$ strengths into the skin-mode, $pn$-mode
and interference fractions,
we show that the ratio of the skin-mode strength to the full strength
may be regarded as a generic function of the excitation energy,
insensitive to nuclides and effective interactions,
particularly beyond Ni.
}
\maketitle
\section{Introduction}
\label{sec:intro}
By recent experiments,
sizable $E1$ strengths have been observed at low excitation energy
in a number of $N>Z$ nuclei,
and are called pygmy dipole resonance (PDR)~\cite{ref:SAZ13}.
Low-energy $E1$ strengths have been predicted in many $N>Z$ nuclei
by a systematic calculation as well~\cite{ref:INY09}.
However, their character has not yet been established.
Although oscillation of the neutron skin against the core (skin mode)
has been argued in connection to the PDR,
there remain other possibilities,
\textit{e.g.} fragmentation of the proton-neutron oscillation ($pn$ mode)
whose dominant part forms the giant dipole resonance (GDR).
As low-energy $E1$ strengths may greatly influence $(n,\gamma)$ reaction rates
under astrophysical environment,
it is significant to comprehend their character
including their energy- and nucleus-dependence.
Moreover, the skin-mode strengths could be correlated
to the slope parameter of the nuclear symmetry energy $L$,
which attracts interest in relevance to structure of neutron stars.

To investigate characters of the low-energy $E1$ excitations,
we have analyzed transition densities obtained from the HF+RPA calculations
in the doubly-magic nuclei~\cite{ref:NIS13}.
By decomposing the transition matrix elements
into the skin-mode and $pn$-mode fractions
via the transition densities,
energy-dependence of the skin-mode fraction has been argued.
Here we extensively study energy-dependence of the skin-mode fraction.

\section{Decomposition of $E1$ strengths}
\label{sec:decomp}
\begin{figure}
% Use the relevant command for your figure-insertion program
% to insert the figure file.
\centering
~\vspace{-1.5cm}\\
\hspace*{1.5cm}\includegraphics[width=8cm]{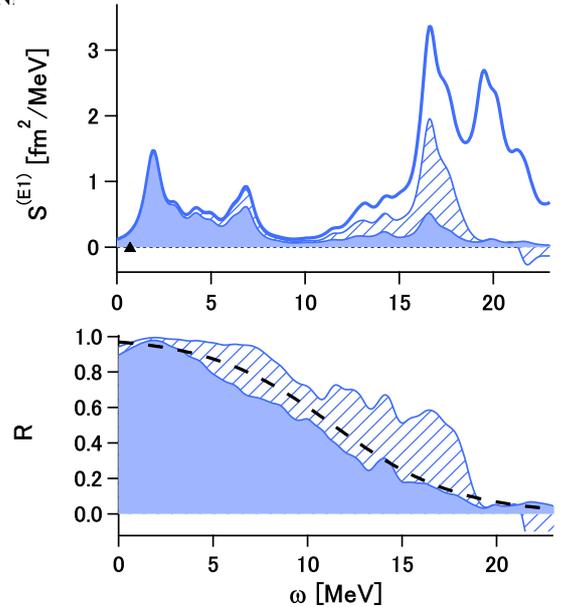}\vspace*{1cm}\\
\caption{$E1$ strength function $S^{(E1)}_\mathrm{mode}(\omega)$ (upper panel)
and ratio $R_\mathrm{mode}(\omega)$ in $^{86}$Ni,
by the HF+RPA calculation with D1S.
Blue-shaded and hatched areas present
the skin mode and the interference contributions, respectively.
Blue thick solid line in the upper panel
gives the full $E1$ strength $S^{(E1)}(\omega)$.
Black triangle attached to the horizontal axis
indicates the neutron threshold in the HF calculation.
Dashed line in the lower panel is $R_\mathrm{skin}(\omega)$
of Eq.~(\ref{eq:Rfit}) with the fitted $\omega_\mathrm{cr}$ and $D_\mathrm{cr}$.}
\label{fig:Sfn}       % Give a unique label
\end{figure}

We have proposed a decomposition method
of the low-energy $E1$ transition matrix elements
into the $pn$ mode and the skin mode via the transition densities,
so that their mixing could be handled in a straightforward manner.
With the proton and neutron transition densities of the excitation
to the $1^-$ state $|\alpha\rangle$,
\begin{equation}
 \delta\rho_{\tau_z}^{(\lambda=1)}(r;\alpha)
  = \big\langle\alpha\big|\sum_{i\in\tau_z} \delta(\mathbf{r}-\mathbf{r}_i)\,
  r_i Y^{(1)}(\hat{\mathbf{r}}_i)\big|0\big\rangle\quad(\tau_z=p,n)\,,
\end{equation}
the $E1$ transition density is given by
\begin{equation}
 \delta\rho^{(E1)}(r;\alpha) = \frac{N}{A}\delta\rho_p^{(\lambda=1)}(r;\alpha)
  -\frac{Z}{A}\delta\rho_n^{(\lambda=1)}(r;\alpha)\,.
 \label{eq:trdns1}
\end{equation}
Depending on the position,
the $E1$ transition density is classified into $\delta\rho_{pn}^{(E1)}$
and $\delta\rho_\mathrm{skin}^{(E1)}$;
if $\delta\rho_p^{(\lambda=1)}/\delta\rho_n^{(\lambda=1)} > -\lambda_s$
with $0<\lambda_s\ll 1$, we take
\begin{equation} \delta\rho_{pn}^{(E1)}(r;\alpha) = 0\,,\quad
 \delta\rho_\mathrm{skin}^{(E1)}(r;\alpha) = \delta\rho^{(E1)}(r;\alpha)\,,
\label{eq:criterion1}\end{equation}
and otherwise,
\begin{equation} \delta\rho_{pn}^{(E1)}(r;\alpha) = \delta\rho^{(E1)}(r;\alpha)\,,
 \quad \delta\rho_\mathrm{skin}^{(E1)}(r;\alpha) = 0\,.
\label{eq:criterion2}\end{equation}
We adopt $\lambda_s=0.05$ below.
The $pn$-mode and skin-mode matrix elements can be obtained
by integrating the transition densities.
The $E1$ strength is then decomposed into the $pn$-mode, skin-mode
and interference contributions for individual $\alpha$.
The corresponding strength function is denoted
by $S^{(E1)}_\mathrm{mode}(\omega)$
(`$\mathrm{mode}$'\,$=$\,`$pn$', `$\mathrm{skin}$' or `$\mathrm{intf}$').
The ratio of $S^{(E1)}_\mathrm{mode}$ to the full strength
%$S^{(E1)}=S^{(E1)}_{pn}+S^{(E1)}_{\mathrm{skin}}+S^{(E1)}_{\mathrm{intf}}$
$S^{(E1)}=\sum_\mathrm{mode} S^{(E1)}_{\mathrm{mode}}$
at the excitation energy $\omega$ is denoted by $R_\mathrm{mode}(\omega)$.

\section{Energy-dependence of skin-mode fraction}
\label{sec:E-dep}
\begin{figure}
% Use the relevant command for your figure-insertion program
% to insert the figure file.
\centering
~\vspace{-1.5cm}\\
\hspace*{1.5cm}\includegraphics[width=8cm]{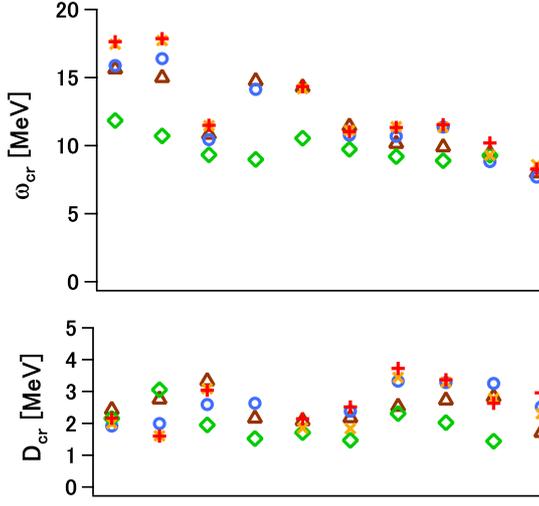}\vspace*{1cm}\\
\caption{Fitted values of $\omega_\mathrm{cr}$ and $D_\mathrm{cr}$,
for $^{22,24}$O, $^{52,60,70}$Ca, $^{68,84,86}$Ni, $^{132}$Sn and $^{208}$Pb
from left to right.
Each symbol represents the effective interaction used in the HF+RPA calculation;
SkI2 (green diamonds), D1S (blue circles), D1M (brown triangles),
M3Y-P6 (orange crosses) and M3Y-P7 (red pluses).}
\label{fig:Rfit}       % Give a unique label
\end{figure}

Figure~\ref{fig:Sfn} displays $S^{(E1)}_{\mathrm{mode}}(\omega)$
and $R_{\mathrm{mode}}(\omega)$ in $^{86}$Ni,
obtained by the HF+RPA calculation with the D1S interaction~\cite{ref:D1S}.
Crossover behavior of the $E1$ excitations has been found,
from the skin mode at low energy to the $pn$ mode at higher energy.
By calculations in a number of spherical $N>Z$ nuclei
with several effective interactions,
the ratio of the skin-mode strength to the full strength $R_\mathrm{skin}$
turns out to have generic energy-dependence,
insensitive to nuclide and to effective interactions
in the energy region of the crossover~\cite{ref:NIS13}.
In order to view this feature more clearly,
we assume a model for $R_\mathrm{skin}(\omega)$,
\begin{equation}
 R_\mathrm{skin}(\omega)
 = \Big[1+\exp\big((\omega-\omega_\mathrm{cr})/D_\mathrm{cr}\big)\Big]^{-1}\,,
 \label{eq:Rfit}
\end{equation}
and adjust the parameters $\omega_\mathrm{cr}$ and $D_\mathrm{cr}$
in individual nuclides for individual interactions
so as to minimize deviation of $R_\mathrm{skin}\cdot S^{(E1)}$
from $S^{(E1)}_{\mathrm{skin}}$.
Degree of fitting is illustrated in the lower panel of Fig.~\ref{fig:Sfn}.
The fitted values of $\omega_\mathrm{cr}$ and $D_\mathrm{cr}$
are depicted in Fig.~\ref{fig:Rfit},
for $^{22,24}$O, $^{52,60,70}$Ca, $^{68,84,86}$Ni, $^{132}$Sn and $^{208}$Pb
with the interactions SkI2~\cite{ref:SkI}, D1S~\cite{ref:D1S},
D1M~\cite{ref:D1M}, M3Y-P6 and M3Y-P7~\cite{ref:Nak13}.
Then $\omega_\mathrm{cr}$ and $D_\mathrm{cr}$ represent
how $R_\mathrm{skin}(\omega)$ varies
according to nuclides and effective interactions.

It is confirmed, as pointed out in Ref.~\cite{ref:NIS13},
that $R_\mathrm{skin}(\omega)$ is insensitive
to nuclides and effective interactions.
In particular, the extracted values of $\omega_\mathrm{cr}$
is quite stable from $^{68}$Ni to $^{208}$Pb,
surprisingly insensitive to the effective interactions,
although there is certain fluctuation in O and Ca.
Fluctuation of the extracted values of $D_\mathrm{cr}$
is not significantly large.
This generic nature of the skin-mode ratio
may be helpful in extracting skin-mode strengths from measurements.

\section{Summary}
\label{sec:summary}
We have extensively investigated characters of the low-energy $E1$ strengths
in $N>Z$ nuclei by the HF+RPA calculations.
Confirming the crossover behavior from the skin mode at low energy
to the $pn$ mode at higher energy,
we apply a method decomposing the $E1$ strengths
into the skin-mode, $pn$-mode and interference fractions,
via the transition densities.
In Ref.~\cite{ref:NIS13}
the ratio of the skin-mode strength to the full strength was suggested
to be a generic function of the excitation energy,
insensitive to nuclides and effective interactions.
By fitting parameters of a model function for the ratio,
this insensitivity is further clarified, particularly beyond Ni.

~\\
This work is financially supported
as Grant-in-Aid for Scientific Research No.~24105008 by MEXT,
and No.~25400245 by JSPS.

%
% BibTeX or Biber users please use (the style is already called in the class, ensure that the "woc.bst" style is in your local directory)
% \bibliography{name or your bibliography database}
%
% Non-BibTeX users please use
%

\end{document}